\documentclass[11pt,preprintnumbers,amsmath,amssymb]{revtex4}
\usepackage[colorlinks=true,urlcolor=blue,linkcolor=blue]{hyperref}
\usepackage{graphicx}
\usepackage{epsfig}
\usepackage{bm}

\usepackage{epsfig}
\usepackage{bm}

\makeatletter

\newcommand{\s}{\sigma}
\newcommand{\la}{\lambda}

\newcommand{\vp}{\vec{\pi}}

\begin{document}

\title{Evolutionary intermittency and the QCD critical point}

\author{N.~G.~Antoniou}
\author{F.~K.~Diakonos}
\author{E.~N.~Saridakis}
\email{msaridak@phys.uoa.gr} \affiliation{Department of Physics,
University of Athens, GR-15771 Athens, Greece}


\begin{abstract}

We investigate the dynamics of the critical isoscalar condensate,
formed during heavy-ion collisions. Our analysis is based on a
simplified model where the sigma and the pions are the only
degrees of freedom. In field description, both in physical and
momentum space, we find that the freeze-out profile presents a
structure which reveals clear traces of the critical fluctuations
in the sigma-component. In particle representation, using
Monte-Carlo simulations and factorial moment analysis, we show
that signatures of the initial criticality survive at the detected
pions. We propose the distribution of suitably defined
intermittency indices, incorporating dynamical effects due to
sigma-pion interaction, as the basic observable for the
exploration of critical fluctuations in heavy-ion collision
experiments.

\end{abstract}

\maketitle

\section{Introduction}

Experiments of a new generation, with relativistic nuclei, at RHIC
and SPS are currently under consideration with the aim to
intensify the search for the existence and location of the QCD
critical point in the phase diagram of strongly interacting matter
\cite{RIKEN,Antoncern}. Important developments in lattice QCD
\cite{karsch} and studies of hadronic matter at high temperatures
\cite{Antoniou2003} suggest that the QCD critical endpoint is
likely to be located within reach at SPS energies. A decisive
observation, in these experiments, associated with the development
of a second-order phase transition at the critical endpoint of QCD
matter is the establishment of power laws in momentum space
(self-similarity) in close analogy to the phenomenon of critical
opalescence in QED matter \cite{lesne}. These power laws reflect
the fractal geometry in real space, and a characteristic index of
the critical behavior is the fractal mass dimension $d_{f2}$ which
measures the strength of the order-parameter fluctuations within
the universality class of critical QCD
\cite{Antoniou2001,reconstruction}.

The physics underlying the endpoint singularity in the QCD phase
diagram is associated with the phenomenon of chiral phase
transition, a fundamental property of strong interactions in the
limit of zero quark masses. In this case and for a given chemical
potential $\mu_B$ there exists a critical temperature $T_{cr}$
above which chiral symmetry is restored and as temperature
decreases below $T_{cr}$, the system enters into the chirally
broken phase of observable hadrons. It is believed that, for two
flavors and zero quark masses, there is a first-order phase
transition line on the $(T,\mu_B)$ surface at large $\mu_B$
\cite{RW1}. This line ends at a tricritical point beyond which the
phase transitions become of second order. In the case of real QCD
with non-zero quark masses, chiral symmetry is broken explicitly
and the first-order line ends at a critical point beyond which the
second-order transitions are replaced by analytical crossovers
\cite{bergraja99}.

The order parameter of this transition is the chiral field
$\Phi=(\s,\vp)$ formed by the scalar, isoscalar field $\s$
together with the pseudoscalar, isovector field
$\vp=(\pi^+,\pi^0,\pi^-)$. Both fields are massless at the
tricritical point and when the symmetry is restored at high
temperatures, their expectation value vanishes,
$\langle\vec{\pi}\rangle=\langle\s\rangle=0$. However, in the
presence of an explicit symmetry breaking mechanism (non-zero
quark masses) the pion and sigma fields are disentangled at the
level of the order parameter of the QCD critical point, which is
now formed by the sigma field alone. The expectation value of the
$\s$-field remains small but not zero near the critical point so
that the chiral symmetry is never completely restored. The
valuable observables in this case are associated with the
fluctuations of the $\s$-field, $(\delta\s)^2\simeq
\langle\s^2\rangle$, and they incorporate, in principle, the
singular behavior of baryon-number susceptibility and in
particular the power-law behavior of the $\s$-field correlator
\cite{hatta03}.

The aim of this work is to study the evolution of critical
correlations during the development of a collision of heavy ions
($A+A$), in the neighborhood of the QCD critical point. In the
initial state we assume that the system has reached the critical
point in thermal equilibrium so that the $\s$-field fluctuations
are described by the 3D Ising critical exponent $\delta$ and in
particular by the fractal dimension
$d_{f2}=\frac{3(\delta-1)}{\delta+1}$ ($\delta\simeq5$). The
crucial question from the observational point of view is whether
traces of the initial criticality can survive in the freeze-out
regime, which follows the equilibration stage, and leave their
imprints at the detectors. In \cite{manosqcd} we showed that this
is indeed the case for a small class of events. However the
treatment presented in \cite{manosqcd} has two important
disadvantages: (i) the analysis of the fluctuations has been
performed in the experimentally inaccessible configuration space
and (ii) the determination of the small class of events carrying
critical characteristics is by itself a non trivial task. In the
present investigation we substantially improve our approach in
this direction, achieving at the same time two main goals: (i)
better adaptation to experimental requirements, by considering the
evolution of the geometry of the critical condensate through
intermittency analysis in transverse momentum space and (ii)
definition of suitable observables which allow for a quantitative
description of the deformation of the critical system by
considering all the events in the ensemble. To proceed in this
direction we introduce the concept of evolutionary intermittency
in transverse momentum space, which is detectable in heavy-ion
collision experiments through factorial moment analysis in an
event-by-event basis. Independently from the duration of the
freeze-out process, the final profile of the deformed system
presents a structure that incorporates all its preceding dynamical
history. For its analysis we investigate the distribution of the
intermittency indices calculated during the evolution towards the
final state. It is shown that this distribution is sensitive to
the initial critical state offering a, free from statistical
restrictions, tool to explore critical fluctuations.

The dynamics of the system is fixed by a two-field Lagrangian,
$\mathcal{L}(\s,\vp)$, together with appropriate initial
conditions. The out-of-equilibrium phenomena are generated by the
exchange of energy between the $\s$-field and the environment
which consists of massive pions initially in thermal equilibrium.
We adopt the picture of a rapid expansion (quench) which is a
realistic possibility in this experimental framework. We study the
out-of-equilibrium evolution of the $\s$-field and we impose its
decay to pions according to the characteristics of the
$\s\rightarrow\pi^{+}\pi^{-}$ procedure. Since in every
collisional event the $\s$-particles decay independently, time
integrations must be taken into account. The freeze-out profile
obtained in this way is analyzed both in the field description and
particle representation. In order to acquire the particle picture,
which is the one obtained in experiments, we use Monte-Carlo
simulations of the aforementioned procedure, and the fractal
measures are quantified through factorial moment analysis
\cite{Bialas88,kittel95}.

In section \ref{model} the formulation of the problem and in
particular the equations of motion and the initial conditions are
presented. In section \ref{physmom} numerical solutions are given
in the context of field description in physical and momentum space
whereas in section \ref{modecarlo} we present numerical results in
particle framework. Finally, in \ref{Conclusions} our results and
conclusions are summarized.

\section{The model}
\label{model}

In our approach we assume an initial critical state of the system
in thermal equilibrium, disturbed by a two-field potential
$V(\s,\vp)$, in an effective description inspired by the chiral
theory of strong interactions \cite{GL,RW}.
The Lagrangian density is
\begin{equation}
\mathcal{L}=\frac{1}{2}(\partial_\mu\s\partial^\mu\s
+\partial_\mu\vp\partial^\mu\vp)-V(\s,\vp) \label{lagr}
\end{equation}
with the potential
\begin{equation}
V(\s,\vp)=\frac{\la^2}{4}(\s^2+\vp^2-v_0^2)^2
+\frac{m_{\pi}^2}{2}\vp^2, \label{pot}
\end{equation}
where $\s=\s(\vec{x},t)$ and $\vp=\vp(\vec{x},t)$. The potential
has the usual $\s$-model form plus a term which breaks the
symmetry along the $\s$-direction. With the addition of the mass
term for the pion field, we ensure that it has a constant mass
equal to $m_\pi$. Finally, the constant terms in (\ref{pot}) shift
the value of the potential at the minimum to zero. We fix the
parameters of the Lagrangian using the phenomenological values
$m_\pi$$\approx139\,$ MeV, $v_0\approx87.4\,$ MeV and
$\la^2\approx 20$ \cite{Gavin93}.

The equations of motion resulting from (\ref{lagr}) are:
\begin{eqnarray}
\ddot{\s}-\nabla^2\s+\la^2(\s^2+\vp^2-v_0^2)\s=v_0m_\pi^2\nonumber\\
\ddot{\vp}-\nabla^2\vp+\la^2(\s^2+\vp^2-v_0^2)\vp+m_\pi^2\vp=0\label{eom1},
\end{eqnarray}
where $\vp^2=(\pi^+)^2+(\pi^0)^2+(\pi^-)^2$.

Using a constant value $v_0$ in eq. (\ref{pot}) implies a
non-vanishing mass (finite correlation length) for the $\s$-field,
$m_\s=\sqrt{2\lambda^2 v_0^2}$, in contradiction with its critical
profile. This inconsistency is restored if we assume a finite-time
mechanism instead of the instant quench i.e. the instantaneous
formation of the potential (\ref{pot}) at $t=0$. The simplest
model which leaves the equations of motion unaffected, used also
in cosmological phase transitions \cite{Rivers}, is the so-called
linear quench \cite{SRS2,manosqcd}. It assumes that the minimum
$v$ of the potential increases linearly with time, starting from
zero and ending at the zero temperature value $v_0\approx87.4\,$
MeV after a time interval $\tau$ which is the quench duration:
\begin{equation}
v(t)=\left\{ \begin{array}{cc} v_0 t/\tau & \text{\ \ \ \ \ \ \ \ \ \ for \ \ \ \ \
$t\leq\tau$}\\
v_0  & \text{\ \ \ \ \ \ \ \ \ \   for \ \ \ \ \ $t>\tau$.}
\end{array}\right.
\end{equation}
This assumption leads naturally to a known evolution of
$m_\sigma(t)=\partial^2V(\s,\vp)/\partial\s^2|_{\sigma=\sigma_{min},\vp=\vp_{min}}$:
\begin{equation}
m_\sigma(t)=\left\{ \begin{array}{cc} \sqrt{2\lambda^2}\frac{v_0t}{\tau} & \text{\ \ \ \ \ \ \
\ \ \ for \ \ \ \ \ $t\leq\tau$}\\
\sqrt{2\lambda^2}v_0 & \text{\ \ \ \ \ \ \ \ \ \ for \ \ \ \ \
$t>\tau$.} \end{array}\right.
 \label{sigmass}
\end{equation}
Thus, $m_\s$ presents the expected behavior, i.e. it is
 $m_\s=0$ at $t=0$ (critical $\s$-field) and, with increasing time,
 $m_\s$ approaches and acquires its zero
temperature value $m_{\s,0}=\sqrt{2} v_0 \lambda$.

We investigate the evolution of the above system using initial
field configurations dictated by the onset of the critical
behavior. In this case we expect that the $\s$-field, being the
order parameter, will possess critical fluctuations, whereas the
$\pi$-fields are thermal, and the entire system will be in
thermodynamical and chemical equilibrium. Obviously, the
subsequent evolution, determined by eqs.~(\ref{eom1}), will
generate strong deviations from equilibrium. The question is
whether the initial critical profile (presenting the critical
point information) survives for times large enough in order to
leave imprints at the freezeout. Before going on with the detailed
study of the dynamics, we first refer to the generation of an
ensemble of critical field configurations (initial conditions),
and we provide the necessary tools for its quantitative
description.

The detailed procedure for the production of the critical ensemble
is given in \cite{manosphysica}. Interpreting the square of the
$\s$-field as local density, we use the effective action at the
critical point \cite{Tsypin}:
\begin{equation}
\Gamma[\sigma]=\int_R d^Dx \{ \frac{1}{2} (\nabla \sigma)^2 + g
\sigma^{\delta+1}  \}, \label{effact}
\end{equation}
(where $\delta$ is the isothermal critical exponent of the
corresponding universality class and $g$ is an associated
effective coupling) in order to weight the sum over field
configurations in the partition function. This sum is saturated
through saddle point solutions which consist approximately of
piecewise constant configurations extended over domains of
variable size \cite{Antoniou98}. In this way we record a large
number of statistically independent configurations possessing
critical characteristics, and the corresponding behavior is
described by a fractal measure demonstrated in the power-law
dependence of the mean "mass" $m_{2x}(R)$ on the distance $R$
around a point $\vec{x}_0$ defined by:
\begin{equation}
m_{2x}(R)=\int_{R} |\sigma^2(\vec{x})\sigma^2(\vec{x}_0)|\,
d^Dx\,d^Dx_0\sim R^{d_{f2x}}, \label{masspower}
\end{equation}
for every  $\vec{x}_0$. $d_{f2x}$ is the fractal mass dimension of
a specific configuration \cite{Mandel83,Falconer}, and the
subscript 2 marks the use of $\s^2$ for its definition, while the
subscript $x$ indicates the calculation in physical space.
$d_{f2x}$ is related to space dimensionality D and isothermal
critical exponent $\delta$ through:
\begin{equation}
d_{f2x}=\frac{D(\delta-1)}{\delta+1}. \label{fracdim}
\end{equation}
For the 3D Ising universality class, $g\approx2$, $D=3$,
$\delta\approx5$, therefore $d_{f2x \mid D=3}\approx2$. The 3D
Ising description has to be adapted to the physical framework of
colliding nuclei. The considered system may be affected by
non-equilibrium processes along the collision axis, and therefore
the picture of statistical equilibration is valid safely only for
the transverse 2D section. Moreover, the relativistic nature of
the longitudinal motion needs an extra assumption (Feynman-Wilson
fluid hypothesis \cite{Feynman}) in order to formulate global
thermal equilibrium in terms of space-time rapidity. It is then
natural to consider the 2D projection of the original 3D Ising
system as a description of the fireball in the transverse space.
Thus, assuming a cylindrically symmetric fractal, its 2D
projection has a mass dimension $d_{f2x \mid 2D \text{\
proj}}\equiv d_{f2x}\approx4/3$, and the $\sigma$-field depends
only on the transverse coordinates. We produce an ensemble of
$10^4$ $\s$-configurations, each one possessing its own fractal
mass dimension, and these $d_{f2x}$'s satisfy (\ref{masspower})
and (\ref{fracdim}) within an error of $2\%$.

Equivalently, we can define the fractal mass dimension in
(transverse) $k$-space  as:
\begin{equation}
m_{2k}(K)=\int_{K}
|\tilde{\sigma}_{2}(\vec{k})\tilde{\sigma}_{2}(\vec{k}_0)|\,
d^2k\,d^2k_0\sim K^{d_{f2k}}, \label{masspowerk}
\end{equation}
where $\tilde{\sigma}_{2}(\vec{k})$ is the 2D Fourier
transformation of $\sigma^2(\vec{x})$, and $K$ is the ``distance"
around a point $\vec{k}_0$ in the transverse momentum space.
Obviously, $d_{f2x}$ and $d_{f2k}$ are related through:
\begin{equation}
d_{f2k}=2-d_{f2x}\label{dfkdfx}.
\end{equation}
Performing numerically the aforementioned Fourier transformation,
we result in a critical ensemble in $k$-space  satisfying
(\ref{masspowerk}), whereas (\ref{dfkdfx}) holds within an error
of less than $10^{-2}$ per cent.

Having established the production algorithm for the critical
ensemble of the $\s$-field, and the necessary measures for its
quantitative description in physical and $k$-space, we can proceed
to its time evolution. As we have already mentioned in the
introduction, we are interested in investigating the evolution of
the system determined by equations (\ref{eom1}), using as initial
conditions an ensemble of $10^4$ independent $\s$-configurations
on the lattice possessing critical characteristics, and $10^4$
configurations, for each $\pi$ component, corresponding to an
ideal gas at temperature $T_0\approx140$ MeV, which simulates the
pion environment in a heavy-ion collision. The production of
thermal $\pi$-configurations has been described in
\cite{Cooper,manosqcd}.

\section{Numerical solutions: Physical and $k$-space
description}\label{physmom}

We evolve the critical system according to (\ref{eom1}) and we
determine the field values at any time. The quantitative
deformation of the initial fractal geometry, is embedded in the
evolution of $m_{2x}(R)$ (or of $m_{2k}(K)$ in $k$-space) and in
particular in the evolution of the exponent of its power-law
dependence on $R$. We call this exponent $\psi_{x}(t)$ (similarly
$\psi_{k}(t)$ in $k$-space). Initially $\psi_{x}(0)\equiv d_{f2x}$
(and $\psi_{k}(0)\equiv d_{f2k}$), i.e. it is the fractal mass
dimension of the critical system.

As we showed in \cite{manosqcd} $\psi_{x}(t)$ reaches in short time interval the
value of the embedding dimension, which in the present case, due
to the projection into transverse space, is equal to 2, and subsequently
fluctuates becoming almost equal to $\psi_{x}(0)$ at particular
times. Thus, the initial critical profile reappears partially and
quite periodically at times when the spatial mean value of the
$\s$-field returns close to its initial value (which is
$\langle\s(0)\rangle\approx0$ as implied by the critical
behavior). Finally, after some oscillations $\psi_{x}(t)$ relaxes
very close to 2, i.e. the initial fractal geometry is completely
lost. The detailed explanation of this phenomenon is given in
\cite{deterministic}.

\begin{figure}[h]
\begin{center}
\mbox{\epsfig{figure=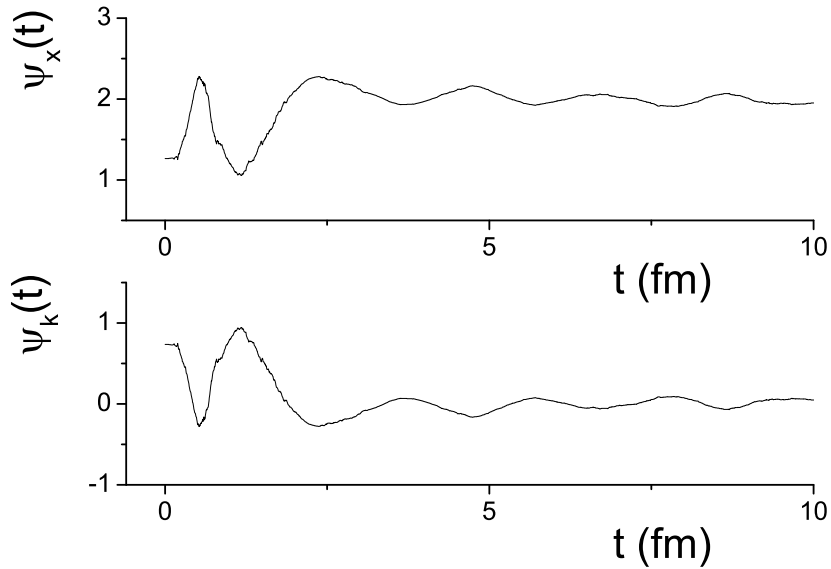,width=11cm,angle=0}} \caption{\it
$\psi_{x}(t)$ (upper graph) and $\psi_{k}(t)$ (lower graph) for
one configuration, for $\tau=5$ fm.} \label{1conf}
 \end{center}
 \end{figure}

In fig.~\ref{1conf} we depict the evolution of $\psi_{x}(t)$ and
$\psi_{k}(t)$ for one configuration of the ensemble, using for the
quench duration the value $\tau=5$ fm, where the calculations are
achieved through the integrals (\ref{masspower}) and
(\ref{masspowerk}) at every time.
We observe the partial re-establishment of the initial critical
profile, and its final complete deformation. Furthermore, we
verify that relation (\ref{dfkdfx}) holds in very good precision
at all times. Lastly, a comment must be made for the overflow of
$\psi_{x}(t)$ above the embedding dimension value 2, and
equivalently for the negatives values of $\psi_{k}(t)$. Actually,
this result is artificial and it arises from small disturbance,
due to finite-size effects, of the linearity of $\ln m_{2x}(R)$
versus $\ln R$, at these time intervals. However, if we proceed to
larger lattices, the exceeding of 2 becomes less strong, as we
have shown in \cite{deterministic}. This is possible for one
event, but it demands extremely huge computational times for the
whole ensemble. Therefore, we remain in the aforementioned
selection, having in mind that slopes larger than 2 (or less than
0) should be consider as being equal to 2 (or 0). Alternatively,
we can avoid this overflow by enforcing a cut on the fit accuracy.
As in this work we are interested in the reappearance of the
initial fractal geometry, our results are trustworthy being not
affected by this behavior.

\begin{figure}[h]
\begin{center}
\mbox{\epsfig{figure=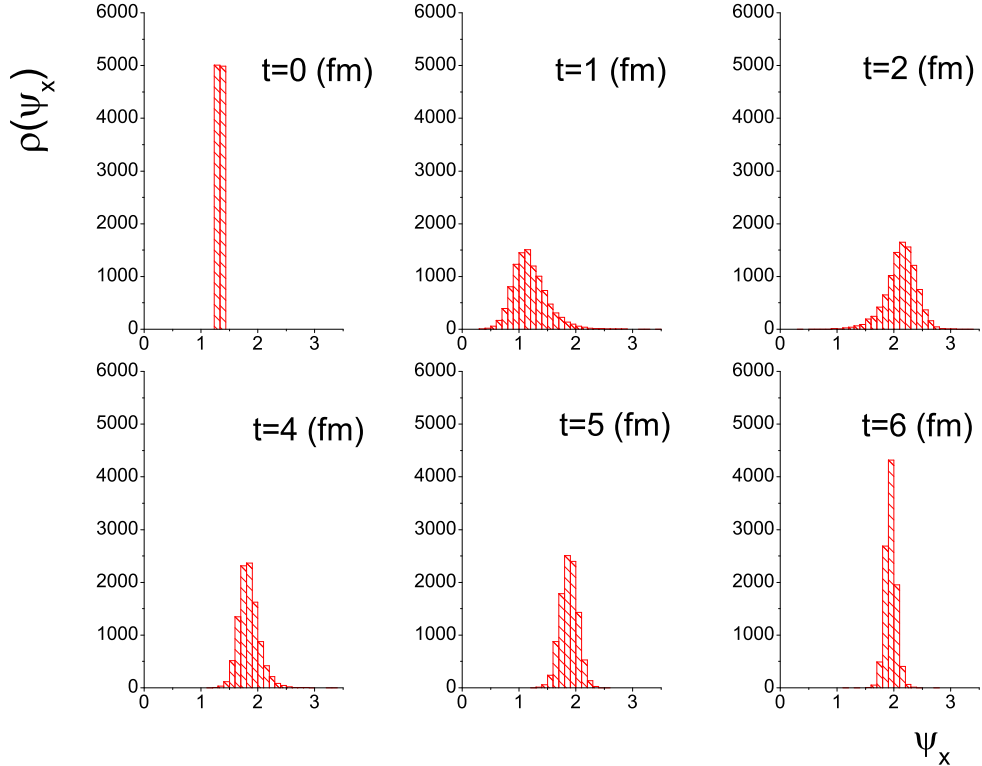,width=13cm,angle=0}} \caption{(color
online)\it Distribution $\rho(\psi_{x})$ for the individual values
of $\psi_{x}(t)$ for each of the $10^4$ independent
$\s$-configurations, for six successive times, for quench duration
$\tau=5$ fm.} \label{dfxproft}
 \end{center}
 \end{figure}

\begin{figure}[!]
\begin{center}
\mbox{\epsfig{figure=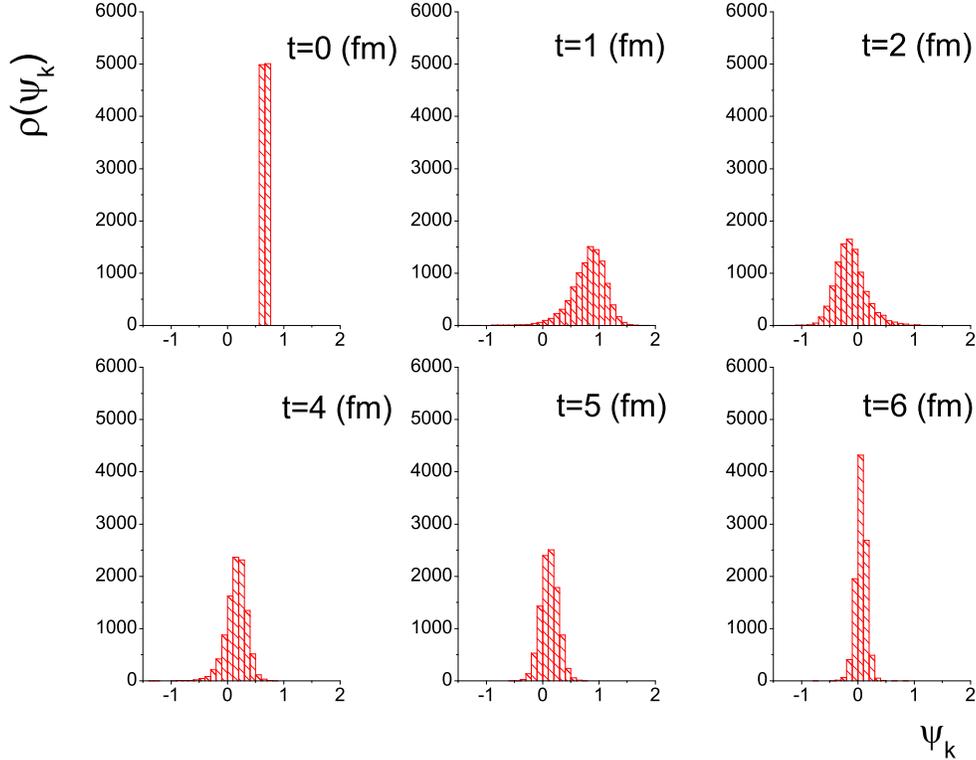,width=13cm,angle=0}} \caption{(color
online)\it Distribution $\rho(\psi_{k})$ for the individual values
of $\psi_{k}(t)$ for each of the $10^4$ independent
$\s$-configurations, for six successive times, for quench duration
$\tau=5$ fm.} \label{dfkproft}
 \end{center}
 \end{figure}

Now, let us examine the evolution of the whole critical ensemble.
To keep the information of the evolution of each individual
configuration, i.e. we perform an event-by-event analysis of the
ensemble evolution. The reason is that, as we showed in
\cite{manosqcd}, the quantitative characteristics of the initial
critical behavior sustaining are more transparent in
fluctuation-related measures rather than in ensemble-averaged
ones. Thus, at each time instant we calculate a histogram for the
distribution $\rho(\psi_{x})$, which corresponds to the individual
values of $\psi_{x}(t)$ for each of the $10^4$ independent
$\s$-configurations.  In fig.~\ref{dfxproft} we depict
$\rho(\psi_{x})$ for six successive times.
As we can see, initially all the $\s$-configurations have
$\psi_{x}(0)\equiv d_{f2x}$ equal to the theoretical value 4/3
(arising from (\ref{fracdim}) with $D=2$, $\delta=5$) within an
error of $2\%$. As time passes, $\psi_{x}(t)$ for each
configuration evolves independently (similarly to
fig.~\ref{1conf}) and both the maximum and the standard deviation
of $\rho(\psi_{x})$ changes. However, as we can clearly observe,
the non-trivial fractal geometry prevails for some time in a
subset of the ensemble, and eventually disappears leading to
$\rho(\psi_{x})$ centered at the value of the embedding dimension
2, with decreasing standard deviation, i.e. all configurations do
gradually lose their initial critical profile. Similarly, we
repeat the same steps for the $k$-space-defined $\psi_{k}(t)$, and
the corresponding evolution of $\rho(\psi_{k})$ is depicted in
fig.~\ref{dfkproft}.
The explanation of the behavior of $\rho(\psi_{k})$ is
straightforward, and as expected the loss of the initial critical
behavior is reflected to an increasingly narrower distribution
$\rho(\psi_{k})$ centered at zero.

\begin{figure}[h]
\begin{center}
\mbox{\epsfig{figure=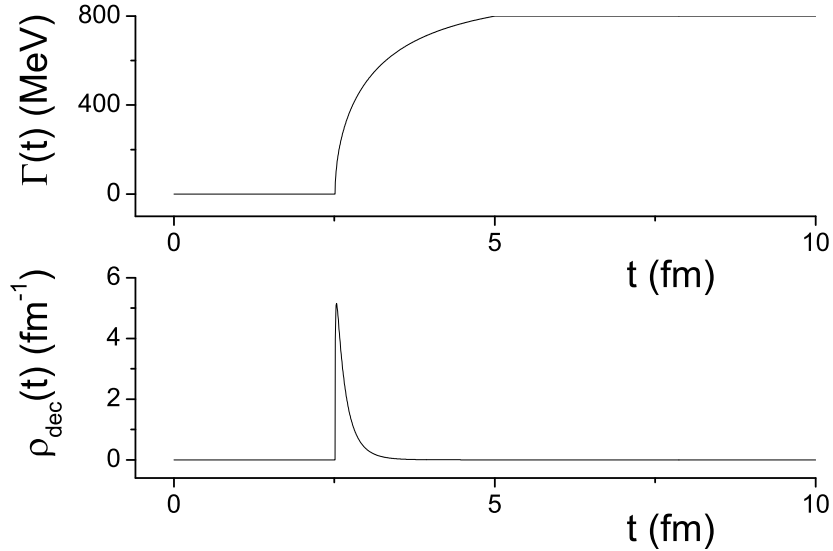,width=11cm,angle=0}} \caption{\it
Decay width $\Gamma(t)$ and $\rho_{dec}(t)$ for $\tau=5~fm$,
$\Gamma_0=800~$MeV and $m_{\s,0}\approx560~$MeV.} \label{Gammarho}
 \end{center}
 \end{figure}

The above evolution of independent configurations is unrealistic
and is only useful in order to demonstrate the fractal geometry
reappearance. In fact, after $m_\sigma(t)$ reaches the threshold
of $2m_\pi$ at time $t_{th}$ which according to (\ref{sigmass})
is:
\begin{equation}
t_{th}=\frac{2m_\pi\tau}{\sqrt{2\lambda^2}v_0}, \label{tthreshold}
\end{equation}
the $\sigma$-particles can decay to pions through
$\sigma\rightarrow\pi^{+}\pi^{-}$. Thus, traces of the initial
criticality can be transferred to the produced pions. In
\cite{manosqcd} we were interested mainly in the qualitative
description of this phenomenon and we showed that signs of the
initial critical behavior can indeed be found in the pion sector. In the present work we proceed to a detailed and more quantitative
analysis. To achieve this, we first have to specify the width of
the $\s$-decay. This is given by the expression:
$\Gamma(t)=\Gamma_0 \sqrt{1-\left(2m_\pi/m_\s(t)\right)^2},$
which is easily treated in our approach since we have assumed a specific (and quite general) form for the quench evolution and consequently for $m_\sigma(t)$ (relation (\ref{sigmass})). Therefore, for the decay width we acquire:
\begin{equation}
\Gamma(t)=\left\{ \begin{array}{cc} 0 & \text{\ \ \ \ \ \ \ \ \ \ for \ \ \ \ \
$t<t_{th}$} \\
\Gamma_0 \sqrt{1-\frac{4m_\pi^2}{2\lambda^2v_0^2}\frac{\tau^2}{t^2}} & \text{\ \ \ \ \ \ \ \
\ \ \ \ \ \ \ \ for \ \ \ \ \ $t_{th}\leq t\leq\tau$}\\
\Gamma_0 \sqrt{1-\frac{4m_\pi^2}{2\lambda^2v_0^2}} & \text{\ \ \ \
\ \ \ \ \ for \ \ \ \ \ $\tau< t$.}\end{array}\right.
\label{Gammat}
\end{equation}
The factor $\Gamma_0$ can be determined by fixing the value of the
$\sigma$-width at zero temperature. Eq.~(\ref{Gammat}) allows us
to determine the probability density $\rho_{dec}(t)$ that a
$\sigma$-particle will decay at $t$, using the definition of
$\Gamma(t)$ in the particle picture: $\frac{d N_{\sigma}}{d
t}=-\Gamma(t)N_{\sigma}$ and neglecting any influence of the
surrounding thermal medium, as well as regeneration effects. The
probability $P(t)$ to decay a sigma particle in the time interval
$[0,t)$, assuming that each sigma decays statistically independent
from the others, is then given as:
$$P(t)=\frac{N_{\sigma}(0)-N_{\sigma}(t)}{N_{\sigma}(0)}=1-e^{-\int_0^t \Gamma(z)~dz}$$
leading to:
$$\rho_{dec}(t)=\frac{d P}{d t}=\Gamma(t) e^{-\int_0^t \Gamma(z)~dz}$$
We finally get the expression:
\begin{equation}
\rho_{dec}(t)=\left\{ \begin{array}{cc} 0 & \text{\ \ \ \ \ \ \ \ \ \ for \ \ \ \ \
$t<t_{th}$} \\
\Gamma_0 \sqrt{1-\frac{4 m_{\pi}^2 \tau^2}{m_{\s,0}^2 t^2}}\,
e^{-\Gamma_0 \left[ \sqrt{t^2-\frac{4
m_{\pi}^2}{m_{\s,0}^2}\tau^2} - \frac{2 m_{\pi}}{m_{\s,0}}\tau
\arccos (\frac{2 m_{\pi} \tau}{m_{\s,0} t}) \right]} & \text{\ \ \
\ \ \ \ \
\ \ \ \ \ \ \ \ for \ \ \ \ \ $t_{th}\leq t\leq\tau$}\\
\Gamma_0 \sqrt{1-\frac{4 m_{\pi}^2}{m_{\s,0}^2}}\, e^{-\Gamma_0
t\sqrt{1-\frac{4 m_{\pi}^2}{m_{\s,0}^2}}  + \frac{2
m_{\pi}}{m_{\s,0}}\Gamma_0 \tau \arccos (\frac{2
m_{\pi}}{m_{\s,0}})} & \text{\ \ \ \ \ \ \ \ \ for \ \ \ \ \
$\tau< t$.} \end{array} \right. \label{rhot}
\end{equation}
Equation (\ref{rhot}) leads to a distribution for the mass of the
decaying $\sigma$'s with a characteristic peak at $m_\sigma^*$,
which for $\tau\rightarrow\infty$ tends to the threshold value
$2m_\pi$
($m_{\sigma}^*\,|_{\tau\rightarrow\infty}\rightarrow2m_\pi$).
Qualitatively, this behavior corresponds to the singular
enhancement of the spectral density of the sigma field in medium
($\rho_{\sigma}(\omega) \sim \left[ 1- \frac{4
m_{\pi}^2}{\omega^2} \right]^{-1/2}$) in the case of partial
symmetry restoration \cite{Kunihiro99}. In practice, using
$\tau=5~fm$ we get $m_\sigma^*\approx282$ MeV (leading to a decay
width $\approx130$ MeV) which is very close to the two-pion
threshold, and the associated $\sigma$-mass distribution
resembles, within our simplified approach, the in-medium spectral
properties of the sigma field at the critical point. In the upper
graph of fig.~\ref{Gammarho} we depict $\Gamma(t)$ for
$\tau=5~fm$, taking $\Gamma_0=800~$MeV and
$m_{\s,0}\approx560~$MeV, while in the lower graph we show the
corresponding $\rho_{dec}(t)$.

Having formulated the quantitative treatment of $\s$-decay we can
proceed to our investigation. The scenario we explore is the
following: We evolve the ensemble of $10^4$ critical
$\s$-configurations in an environment of thermal pions. After
$t_{th}$ each one of them will decay independently according to
$\rho_{dec}(t)$ given by (\ref{rhot}), i.e. each configuration
will decay at each own $t_{dec}$, and this corresponds to a single
event. However, experimentally we get only a freeze-out profile of
events, since there is no available information concerning the
decay time at which the pions, observed at the detectors, were
produced. To acquire this profile we have to dress the evolution
of $\rho(\psi_{x})$ with the decay probability $\rho_{dec}(t)$.
Therefore, the event-by-event freeze-out profile in physical space
will be the integral of $\rho_{\psi_{x}}(t)$ weighted with
$\rho_{dec}(t)$, up to a final time corresponding to the
freeze-out time $t_{frz}$:
\begin{equation}
\rho_{w\psi_{x}}(t_{frz})=\int_0^{t_{frz}}\rho_{\psi_{x}}(t)\rho_{dec}(t)\,dt.\label{rhoconv}
\end{equation}
 Although in general the weighted integral has to be
performed up to infinite freeze-out time, (or up to its
experimentally determined value), due to the exponential decrease
of $\rho_{dec}(t)$ all the profiles with freeze-out
times $t_{frz}$ larger than $t_{*}\approx10 fm$ are practically
identical. Similarly, the corresponding profile in $k$-space will
be the weighted integral $\rho_{w \psi_{k}}(t_{frz})$ of
$\rho_{\psi_{k}}(t)$ with $\rho_{dec}(t)$.

\begin{figure}[h]
\begin{center}
\mbox{\epsfig{figure=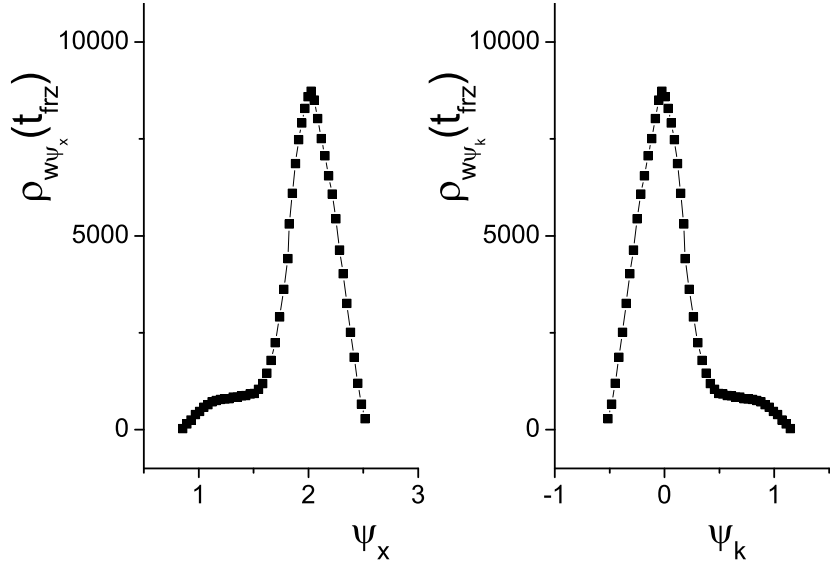,width=11cm,angle=0}} \caption{\it On
the left the freeze-out profile in physical space, given as the
weighted integral $\rho_{w\psi_{x}}(t_{frz})$ of
$\rho_{\psi_{x}}(t)$ with $\rho_{dec}(t)$, for $t_{frz}=10$  fm.
On the right the corresponding profile in $k$-space, given as the
weighted integral $\rho_{w\psi_{k}}(t_{frz})$  of
$\rho_{\psi_{k}}(t)$ with $\rho_{dec}(t)$, for the same freeze-out
time.} \label{wintegralxk}
 \end{center}
 \end{figure}

In fig.~\ref{wintegralxk} we depict this
freeze-out profile, both in physical and $k$-space, for freeze-out
time $t_{frz}=10$ fm (normalized to give the correct
event-multiplicity, i.e. $10^4$).
As we observe, apart from the main peak at 2 in physical space
(the main peak at 0 in $k$-space) which corresponds to complete
loss of the initial critical behavior, there is a secondary
``shoulder'' which is a result of the time history of all the
events that preserved traces of their initial critical profile at
their decay-time $t_{dec}$. This secondary structure is a safe and
robust signature of the initial criticality. Furthermore, it is
also univocal, since starting from conventional (non-critical)
initial conditions it is impossible to acquire even one
non-conventional event at freeze-out, as we have shown in
\cite{manosqcd}.

\section{Numerical solutions: Monte-Carlo simulation}\label{modecarlo}

In the previous section we investigated the evolution of the
critical system, constituted by a large number of individual
events, based on the elaboration of the $\s$ and $\vp$ fields. In
this section we proceed to a more advanced description and refer
to particles, since these are the entities that are detectable.
This process demands a sequence of steps in order to bring our
treatment closer to the conditions present in a heavy-ion collision
experiment.

In our study we use $\s$-configurations which correspond to single
events and we assign to each of them a $\s$-multiplicity of
$\approx140$ $\s$'s. This corresponds to $A+A$ collisions with
intermediate size nuclei assuming the ideal case when the charged
pions observed in the (hypothetical) detector originate
exclusively from the decay of these sigmas and they do not suffer
final state interactions. In fact, the integral: $\int_V
\sigma^2(\vec{x}) d\vec{x}$ is much larger than the integral:
$\int_V \vec{\pi}^2(\vec{x}) d\vec{x}$ for all times. Therefore,
interpreting $\sigma^2$ as the sigma-particle density and
$\vec{\pi}^2$ as the pion density, the above property indicates
that at criticality the multiplicity of the pions in the thermal
environment is much smaller than the corresponding
$\s$-multiplicity. Now, the presence of a large number of
sigma-particles per event can dramatically complicate things since
every $\s$-particle within an event can decay independently, and
there is no consistent formalism for treating the subsequent
evolution of the $\s$-field after its partial decay. However,
since $\rho_{dec}(t)$ has a sharp peak, we assume that all the
$\s$'s within a single event decay at the same $t_{dec}$, which is
of course varying for the different events of the ensemble.

Conclusively we follow this procedure: We evolve each
configuration (individual event) according to (\ref{eom1}). We
determine the corresponding $t_{dec}$ according to
$\rho_{dec}(t)$, and at this specific time we find the 2D Fourier
transformation $\tilde{\sigma}_{2}(\vec{k})$ of the square of the
$\s$-field $\sigma^2(\vec{x})$. Then we produce the momentum
components of each of the 140 particles through a Monte-Carlo
simulation of this momentum profile. Knowing the momenta of the
$\s$-particles we can easily use the kinematics of the decay
$\sigma\rightarrow\pi^{+}\pi^{-}$, in order to determine the
momenta of the produced pions, similarly to the previous work
\cite{Antoniou2001}. Note that our algorithm allows a loss of
information through the secondary decay
$\sigma\rightarrow2\pi^{0}$, and therefore the produced pion pairs
are less than the 140 $\s$'s (they vary between 80 and 100, i.e.
around the two thirds of the $\s$-multiplicity). Finally, we
repeat this procedure for the whole ensemble of events.

In particle representation, the requisite statistical tool for the
quantitative description of the fractal properties is the
factorial moment analysis. We first perform a partition of the
two-dimensional transverse $k$-space into $M^2$ cells, and we
calculate the particle multiplicity $n_m$ inside each cell. Then
we repeat this for all possible cell sizes (between two scales).
The corresponding second factorial moment is defined as:
\begin{equation}
F_2(M)=\frac{\frac{1}{M^2}\sum_{m=1}^{M^2}n_m(n_m-1)}
{\left(\frac{1}{M^2}\sum_{m=1}^{M^2}n_m\right)^2},\label{moment2}
\end{equation}
and for a critical profile, corresponding to isothermal critical
exponent $\delta$, this two-dimensional transverse moment analysis
leads to the power law:
\begin{equation}
F_2(M)\sim(M^2)^{f_2},\label{momentpow}
\end{equation}
with the exponent $f_2$ given by:
\begin{equation}
f_2=\frac{\delta-1}{\delta+1}.\label{momentexp}
\end{equation}
$f_2$ is called ``intermittency index'', since (\ref{momentpow})
determines the self-similarity of the multi-particle state
spectrum, when it is being analyzed in different scales, the
so-called intermittency \cite{Antoniou2001,Bialas88,kittel95}.
Note that since we work in the framework of event-by-event
analysis throughout this study, we define the second factorial
moment for each of the individual configurations, and thus we do
not perform event averaging in (\ref{moment2}) which is usually
taken in the literature. This leads to a slight decrease in the
quality of the fit in $F_2(M)$ vs $M^2$ graph, due to reduced
statistics. However, as we have already mentioned, the
event-by-event analysis reveals rich information about the
fluctuations evolution, which remains hidden in an average
analysis.

There is one last problem that remains to be solved, before the
framework of our approach is completed. Although, given the
$\s$-particle momenta, we can verify relation (\ref{momentpow}) in
large accuracy, the pion analysis presents a complication. In an
event of a heavy-ion collision the majority of the $\s$-particles
decays to a pair of $\pi^{+}$,$\pi^{-}$, while the rest decay to
neutral pions which cannot be detected straightforward. However,
in the detectors we record all the charged pions of an event
simultaneously, and thus we cannot find which pairs of them come
from the same $\s$, in order to reconstruct them and compute the
factorial moments correctly. A more sophisticated reconstruction
procedure is necessary, and it has been elaborated in
\cite{reconstruction}. The basic features are the following: We
first choose a narrow invariant mass window, located just above
the two-pion threshold, to perform the reconstruction of the sigma
field. This is dictated by the form of the $m_\sigma$-distribution
(which for $\tau=5~fm$ is peaked at $m_\sigma^*\approx282$ MeV)
and is compatible with the spectral properties of the critical
sigma-field in medium due to the partial restoration of chiral
symmetry \cite{Kunihiro99}.
 We then combine all $\pi^{+}$ with all $\pi^{-}$ within a
single event making di-pions, and we record the subset of them
that has invariant mass inside the selected window. Obviously,
some of these di-pions correspond to real $\s$'s, but some of them
(usually the majority) consist a combinatorial background
modifying the value of the factorial moments. This background can
be simulated by di-pions reconstructed from mixed events and it is
removed by the substraction of the corresponding moments. The
corrected factorial moments reveal, to a large extent, the
behavior of the $\s$'s before their decay. Moreover, this
substraction eliminates also the thermal pions that are present in
our model and give conventional contribution to the factorial
moments. We point here that in the reconstruction procedure there
is an additional information loss, concerning the initial sigma
state, in the neutral pion channel. Furthermore, performing it
event by event it stiffens the procedure due to the low
statistics. However, in the end we re-acquire an exponent of the
factorial moments that is equal to the expected within an error of
less than $5\%$.

\begin{figure}[h]
\begin{center}
\mbox{\epsfig{figure=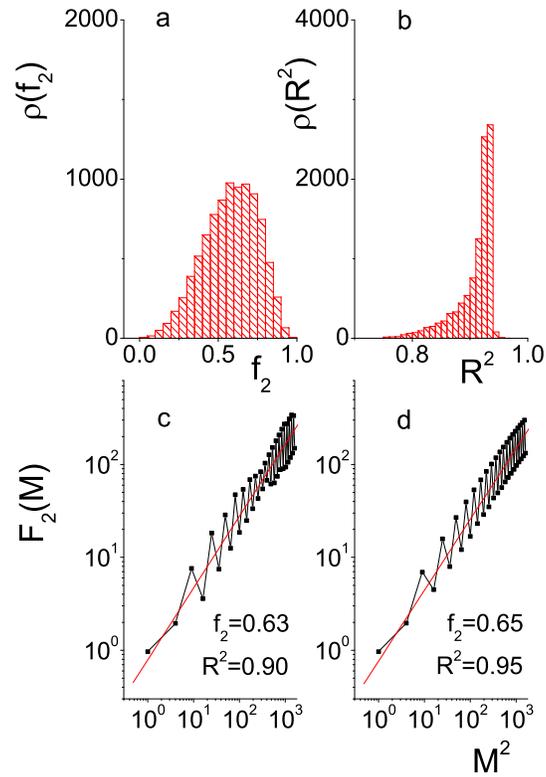,width=7.9cm,angle=0}}
\caption{(color online)\it In graph a) we present the histogram
$\rho(f_2)$ of the exponents $f_2$ of the second factorial moment
for each one of the events, in the case of $\sigma$-particles
produced directly from the critical ensemble corresponding to
$\delta=5$. In graph b) we show the corresponding values of $R^2$
which determine the precision of the fits. In graph c) we depict
the second factorial moment of a single event (with slope less
than one fifth standard deviations away from the histogram
average). In graph d) we present the average second factorial
moment, which is usually used in the literature.} \label{momst0}
 \end{center}
 \end{figure}

Having formulated our method, we can proceed to numerical
simulation, verifying first the validity of our approach at each
step. In order to test the Monte-Carlo procedure and the factorial
moment analysis in the $\s$-sector, we simulate the decay of the
initial critical $\s$-configurations (corresponding to $\delta=5$)
and we perform moment analysis on the produced $\s$-particles. In
subfigure~\ref{momst0}a we depict the second factorial moment for
each one of the events and in subfigure ~\ref{momst0}b we present
the corresponding $R^2$ values which control the quality of the
linear fit. $R^2$ is defined as:
\begin{equation}
R^2=\frac{(<y \tilde{y}>-<y>
<\tilde{y}>)^2}{(<y^2>-<y>^2)(<\tilde{y}^2>-<\tilde{y}>^2)},\nonumber
\end{equation}
where $y$ are the values for a given measure $Y$ while $\tilde{y}$
are the corresponding values of the fitting function.
Each value of these histograms arise from a single event, where
the corresponding second factorial moment plot is like the one
shown in subfigure \ref{momst0}c (here we show an event with slope
that is one fifth standard deviations from the histogram average).
As we can see from the $\rho(f_2)$ histogram, the obtained $f_2$
values in the $\s$-particle sector are distributed around the
theoretical expectation $f_2=2/3$ according to relation
(\ref{momentexp}). The width of the distribution in general
depends on the multiplicity of $\s$-particles per event which in
our case is fixed to 140. Thus, within our treatment the critical
profile of the $\s$-particles, through both Monte-Carlo simulation
and factorial moment analysis, is reproduced in a consistent way.
Lastly, in fig.~\ref{momst0}d we depict the usually calculated
average second factorial moment. Although linearity (described by
$R^2$) is here better and the obtained slope is close to the
expected one, the average calculation has the disadvantage of
information loss due to the replacement of the entire structure of
the $\rho(f_2)$-histogram by a single value. As we see in the
following the overall structure of the intermittency index
histogram proves to be very important and this is the reason why
we insist on the event-by-event analysis. Finally, note here that
the average factorial moment of fig.~\ref{momst0}d is not exactly
the average of the histogram $\rho(f_2)$, since in the former case
the mean is calculated before the sum in relation (\ref{moment2}),
while in the latter we take the average of the slopes.

\begin{figure}[h]
\begin{center}
\mbox{\epsfig{figure=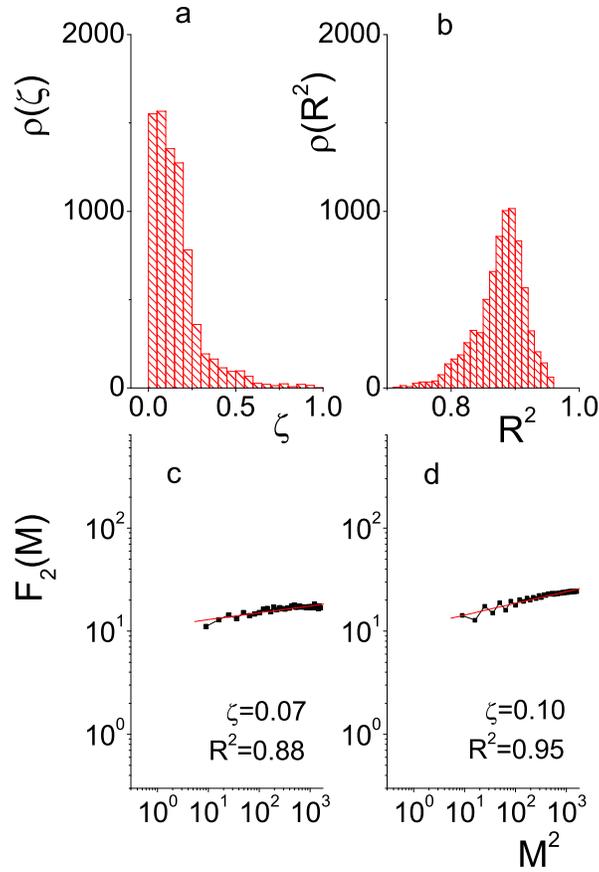,width=7.9cm,angle=0}}
\caption{(color online)\it In graph a) we present the histogram
$\rho(\zeta)$ of the exponents $\zeta$ of second factorial moment,
for all the event population, arising from a conventional moment
analysis, in $\delta=5$ case, for $t_{frz}=10$ fm. In graph b) we
show the corresponding values of $R^2$ which determine the
precision of the fits. In graph c) we depict the second factorial
moment of a single event (with slope less than one third standard
deviations away from the histogram average), for $t_{frz}=10$ fm.
In graph d) we present the average second factorial moment, for
$t_{frz}=10$ fm.} \label{mompinorec}
 \end{center}
 \end{figure}

Having verified our approach at the level of the initial
$\s$-configuration we can proceed to the time evolution. We evolve
the system of $10^4$ critical $\s$-configurations (events) in an
environment of $10^4$ thermal pion-configurations and we let them
decay independently  according to $\rho_{dec}(t)$. From each
configuration we get 140 $\s$-particle momenta through Monte-Carlo
simulation, and using kinematics we perform the decay with a
branching ratio 2:3 to $\pi^{+}$,$\pi^{-}$ pairs. These are
simultaneously detected at the freeze-out (similarly to the
previous section we use a value of $t_{frz}=10$ fm which is by far
satisfactory), prohibiting the identification of the original
$\sigma$'s. Thus, as stated above, due to the combinatorial
background, a simple factorial moment analysis will not be
sufficient to reveal the critical properties of the decaying
system. Indeed, in fig.~\ref{mompinorec} we present the
corresponding four graphs of the moment analysis at the
freeze-out.
In subfigure~\ref{mompinorec}a we show the histogram $\rho(\zeta)$
of the exponent $\zeta$ for the charged pions, which for a fractal
profile is equal to $f_2=2/3$ according to (\ref{momentexp}),  and
in subfig.~\ref{mompinorec}b we give the corresponding $R^2$
values. In subfig.~\ref{mompinorec}c we depict the second
factorial moment for a single event (with slope less than one
third standard deviations away from the histogram average) while
in ~\ref{mompinorec}d we present the average factorial moment for
the whole ensemble of events. As we mentioned, the simultaneous
detection of all the pions suppresses in this sector the signal
associated with the partial restoration of the critical sigma
profile, and reduces the obtained $\zeta$'s values. Thus, in order
to recover the signature of the initial fractal geometry, the
previously described more sophisticated reconstruction procedure
is necessary.

\begin{figure}[h]
\begin{center}
\mbox{\epsfig{figure=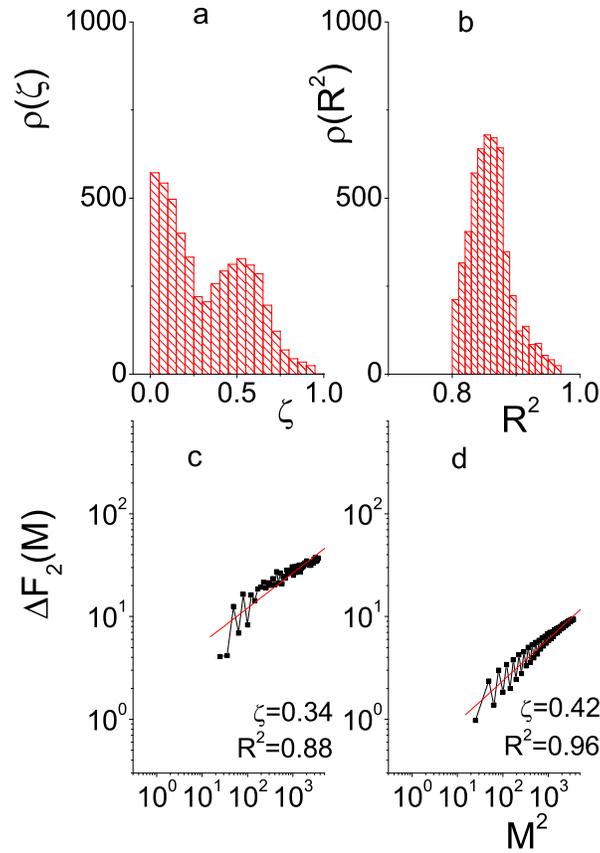,width=7.9cm,angle=0}}
\caption{(color online)\it In graph a) we present the histogram
$\rho(\zeta)$ of the exponents $\zeta$ of second factorial moment,
for all the event population at the freeze-out ($t_{frz}=10$ fm),
performed after the reconstruction procedure with an analysis
window of $10$ MeV in the di-pion invariant mass, in $\delta=5$
case. In graph b) we show the corresponding values of $R^2$ which
quantify the linearity accuracy. In both a) and b) subfigures we
have kept only those reconstructed events which possess $R^2>0.8$.
In graph c) we depict the second factorial moment of a single
reconstructed event (with slope less than one fifth standard
deviations away from the histogram average), for the same
universality class, analysis window and freeze-out time. In graph
d) we present the average second factorial moment for the ensemble
of reconstructed events.} \label{mompirecon}
\end{center}
\end{figure}

In fig.~\ref{mompirecon} we depict the results of the factorial
moment analysis for the pions at the freeze-out, after the
reconstruction step, using an analysis window of $10$ MeV just
above the two-pion threshold $2 m_{\pi}$. This particular choice
is appropriate since it includes the $m_{\pi^+ \pi^-}$ invariant
mass value which corresponds to the maximum of the mass
distribution of the decaying sigmas ($m_\sigma^*\approx282$ MeV).
In addition, in this kinematical window the probabilities of the
appearance of ($\pi^+ \pi^-$)-pairs originating from the decay of
conventional resonances (as for example the $\rho$-meson) is
vanishingly small. Furthermore, performing our analysis so close
to the two-pion threshold has another advantage: at these
invariant-mass values the re-generation of sigmas from the pions
will be small, i.e the generation of a $\sigma$-particle with this
constrained invariant-mass, from pions produced by primary sigma
decay, is statistically rare, which provides a justification for
non-consideration of re-generation effects. Finally, the mean
multiplicity of $\sigma$'s, before reconstruction, within this
window is $\approx60$.
$\Delta F_2(M)$ stands for the corrected factorial moments, which
arise after the subtraction of the moments of the mixed events.
The reconstruction procedure in the event-by-event analysis,
compared to the reconstruction in the whole ensemble which is
widely used in the literature, possesses some additional problems.
Firstly, in a small fraction of events the subtraction of the
moments of the mixed pions leads to negative moments. This is a
sign that for these events, and at the specific cell partition,
the reconstruction procedure did not work and therefore they have
to be excluded. Secondly, another fraction of reconstructed events
possesses positive moment values with slopes (i.e. $\zeta$-values)
smaller that 0 or greater than 1. This is also a result of an
unsuccessful reconstruction and means that their profile is
completely overwhelmed by the background of mixed events. Both
these problem sources have their origin in the low statistics of
the event-by-event analysis, contrary to the ensemble averaged
one. However, for the majority of events the procedure is
successful, and this allows us to acquire the information which
would be hidden in an average analysis.

The time-evolution affects the initial geometry of the critical
system in a two-fold way: Firstly, the power law weakens, i.e.
most of the exponents $\zeta$ are smaller than the critical value
$f_2=2/3$. Secondly, the linearity demanded in  relation
(\ref{momentpow}) is spoiled compared to an ideal fractal case.
Both these features are obvious in fig.~\ref{mompirecon}. Note
that in order to ensure the power-law behavior we cut in the final
analysis the reconstructed events with $R^2<0.8$. These events
either correspond to a partially unsuccessful reconstruction, or
(which is the usual case) their time evolution has significantly
spoiled the initial critical profile. In both cases they are of no
interest for the purposes of this work, which focus on the
survival of criticality. After their elimination we remain with
about three fifths of the initial population.

Although time evolution weakens the power-law strength and spoils
its form, one can clearly observe strong signs of the initial
criticality. That is, the slope $\zeta\approx0.42$ of
fig.~\ref{mompirecon}d is rather large and linearity is
satisfactory. This result is displayed more transparently in
subfigure \ref{mompirecon}a. The peak of $\rho(\zeta)$ at small
$\zeta$ reveals the complete loss of the initial criticality in
the majority of events. However, the structure of $\rho(\zeta)$
and especially its secondary peak is a robust sign that the
initial fractal geometry has been survived in a subset of events.
Quantitatively, we can calculate the average $\zeta$ for those
events of fig.~\ref{mompirecon}a that are on the right of the
minimum that separates the conventional background from the
secondary peak. We find
$\langle\zeta\rangle|_{>\text{min.}}\approx0.58$, which is very
close to the initial value $2/3$ of the critical index. Thus,
evolutionary intermittency, in an event-by-event basis, provides a
clear and unambiguous measure of revealing the signs of the
survived criticality. Finally, we have confirmed that, although
always carrying the information of the initial critical geometry,
$\langle\zeta\rangle|_{>\text{min.}}$ is sensitive to the value of
the isothermal critical exponent $\delta$ (which determines the
universality class). This behavior enhances the robustness of
evolutionary intermittency in the investigation of the QCD
critical point, and furthermore it supports its use in order to
trace the QCD universality class.

\section{Summary and Conclusions}
\label{Conclusions}

In this work we have explored, both in physical and momentum
space, the dynamics of the critical isoscalar condensate which is
expected to be formed near the QCD critical point. For this
purpose we have adapted the $\s$-model Lagrangian in order to
describe correctly the characteristics of the order parameter
associated with the critical endpoint of the QCD phase transition
\cite{RW} and we have developed a scheme to propagate the
corresponding $\s$-configurations in momentum space. The issue is
of primary importance in the search for the existence and location
of the QCD critical point, in experiments with nuclei. At the
phenomenological level these fluctuations are expressed through
the fractal mass dimension of the $\s$-field configurations,
determining the properties of the condensate at criticality
\cite{Antoniou2001,manosqcd}.

In particular, we have considered the geometrical deformation of
the transverse momentum pattern of the critical $\s$'s as a
function of time, in terms of factorial moment analysis. To take
into account dynamical effects related with this deformation we
have introduced the concept of evolutionary intermittency,
quantified by the total distribution of the critical indices
$\zeta$. We have calculated them by investigating the transverse
momentum factorial moments in an event-by-event basis, assuming
that each event is characterized by a different decay time lying
in the freeze-out zone. Thus, evolutionary intermittency,
describing event-by-event dipion pair correlations in transverse
momentum space, is the basic observable in our approach. It
incorporates, in terms of fractal mass dimension values, all the
out-of-equilibrium dynamical history of the critical system. In
this context, the main result of our treatment is the survival of
critical fluctuations in momentum space even if dynamical effects
prevail in the freeze-out era. This survival is exrpessed as a
secondary peak in the distribution of the critical indices
representing the evolutionary intermittency profile of the
considered process.

The description of the critical fluctuations for the isoscalar
condensate within the presented approach is compatible, at least
in an approximate scheme, with (a) the sigma dynamics, (b) the
initial conditions of a second-order phase transition, (c) the
effects in medium and in particular the partial restoration of
chiral symmetry and (d) the requirements of observability in
nuclear collisions. However, we do not pretend to present a
complete picture since the introduced quench scenario does not
allow for the parametrization of the freeze-out state of the
formed critical condensate in terms of chemical potential and
temperature, which are measurable quantities in a heavy-ion
collision experiment. In addition we have neglected collective
transverse flow phenomena. To incorporate these issues, one should
include additional constraints in the dynamics which are beyond
the present study. Last but not least, the assumption of pions
originating exclusively from critical sigmas and not suffering
final-state interactions, constitutes an idealized scenario for
the evolution of the critical system. Nevertheless, despite its
limitations, the treatment of the present work is a realistic step
towards the determination of suitable observables associated with
the critical fluctuations in a heavy-ion collision experiment.
 The obtained result are of interest
especially in front of the forthcoming experiments at RHIC and SPS which are expected to reach this region in the QCD phase diagram.\\

\paragraph*{{\bf{Acknowledgements:}}} The authors acknowledge partial financial support through
the research programs ``Pythagoras'' of the EPEAEK II (European
Union and the Greek Ministry of Education) and ``Kapodistrias'' of
the University of Athens.

\end{document}